\documentclass[structabstract]{aa}
\usepackage{graphicx}
\usepackage{txfonts}
\usepackage{rotating}
\usepackage{lscape}
\usepackage{longtable}
\usepackage{natbib}
\bibpunct{(}{)}{;}{a}{}{,}

\begin{document}

\title{The extended population associated with W40
\thanks{Based on observations collected at the Centro Astron\'omico Hispano en Andaluc\'\i a (CAHA) at Calar Alto, operated jointly by the Junta de Andaluc\'\i a and the Instituto de Astrof\'\i sica de Andaluc\'\i a (CSIC).}
}
\author{F. Comer\'on\inst{1}
\and A.A. Djupvik\inst{2,3}
\and N. Schneider\inst{4}
}
 \institute{
  European Southern Observatory, Karl-Schwarzschild-Str. 2, D-85748 Garching bei M\"unchen, Germany\\
  \email{fcomeron@eso.org}
  \and
  Nordic Optical Telescope, Rambla Jos\'e Ana Fern\'andez P\'erez, 7, E-38711 Breña Baja, Spain
  \and
  Department of Physics and Astronomy, Aarhus University, Ny Munkegade 120, DK-8000 Aarhus C, Denmark
  \and
I. Physik. Institut, University of Cologne, Z\"ulpicher Str. 77, D-50937 Cologne, Germany
  }
%
%
\date{Received; accepted}
\abstract
{W40 is a heavily obscured bipolar HII region projected in the direction of the Aquila Rift and ionized by hot stars in a central, partly embedded cluster. The study of the cluster and its surroundings has been greatly hampered thus far by the strong extinction in the region.}
{We aim to improve the characterization of the W40 central cluster in terms of the census of its members and their spectral classification. We also search for other members of the region outside the central cluster, with particular interest in previously unidentified massive members that may contribute to the energy budget powering the expansion of the HII region.}
{We use the Gaia eDR3 catalog to establish astrometric membership criteria based on the population of the W40 central cluster, reassess the distance of the region, and identify in this way new members, both inside and outside the cluster. We obtain visible spectroscopy in the red spectral region to classify both known and new members, complemented with Gaia and Spitzer photometry to assess the evolutionary status of the stellar population.}
{Based on stars with high quality Gaia astrometry we derive a high-confidence geometric distance to the W40 region of 502~pc~$\pm$~4~pc and confirm the presence of a comoving extended population of stars at the same distance, spreading over the whole projected area of the HII region and beyond. Spectral classifications are presented for 21 members of the W40 region, 10 of them belonging to the central cluster. One of the newly identified B stars in the extended population is clearly interacting with the shell surrounding the HII region, giving rise to a small arc-shaped nebula that traces a bow shock. The infrared excess properties suggest that the extended population is significantly older ($\sim 3$~Myr) than the W40 central cluster ($< 1$~Myr).}
{The area currently occupied by the W40 HII region and its surroundings has a history of star formation extending at least several million years in the past, of which the formation of the W40 central cluster and the subsequent HII region is one of the latest episodes. The newly determined distance suggests that W40 is behind, and physically detached from, a pervasive large dust layer which is some 60~pc foreground to it as determined by previous studies.}


\maketitle

\section{Introduction \label{intro}}

\begin{figure}[ht]
\begin{center}
\hspace{-0.5cm}
\includegraphics [width=8.5cm, angle={0}]{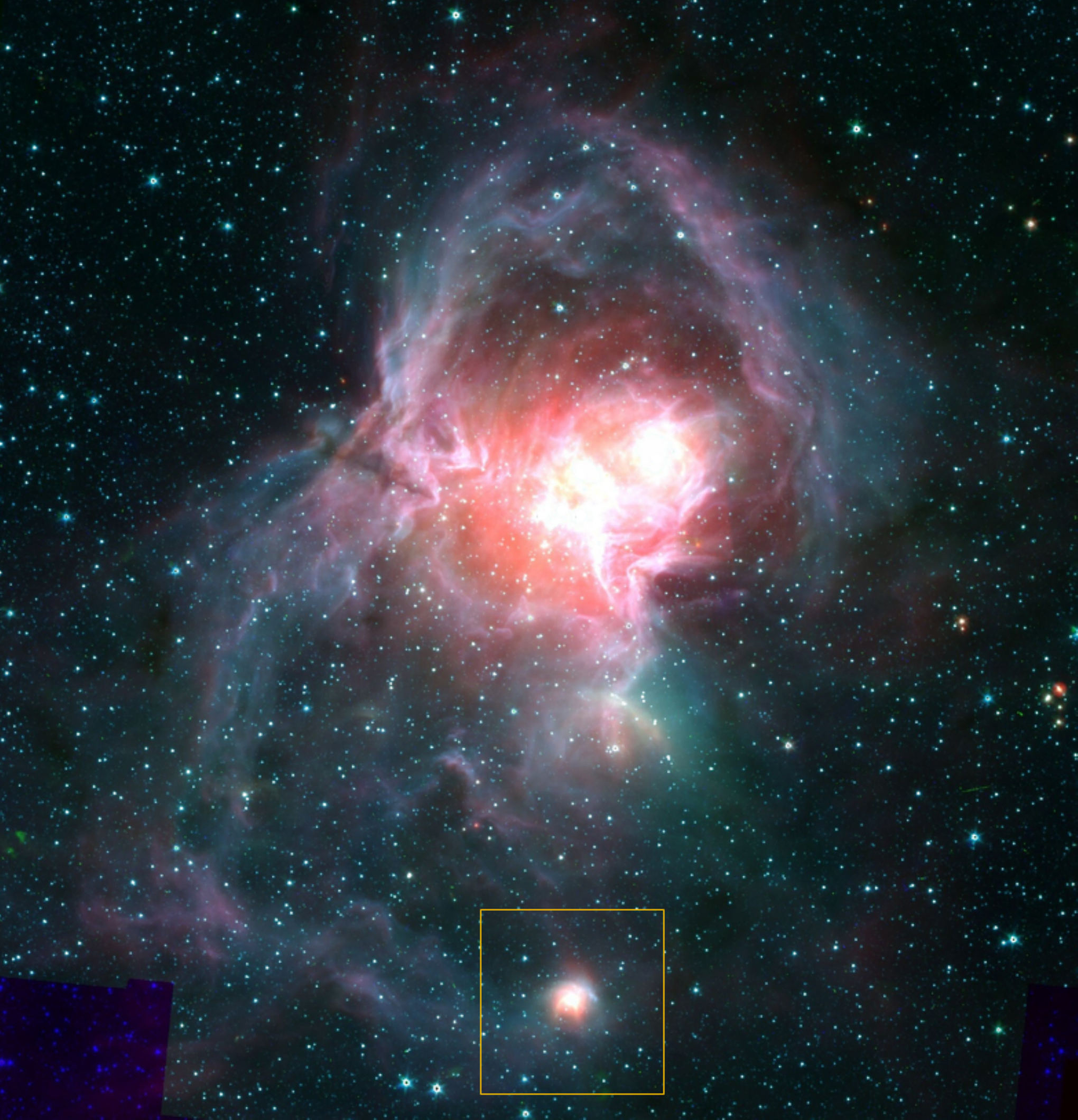}
\caption []{Panoramic view of W40 obtained by the Spitzer Space Observatory. The blue, green and red channels correspond to emission in the 3.6, 4.5, and 24~$\mu$m bands. The field covered is $34'$ across. The yellow rectangle indicates the area displayed in more detail in Figure~\ref{bowshock}.}
\label{W40_Spitzer}
\end{center}
\end{figure}

At a distance comparable to that of the Orion complex, W40 is a relatively nearby HII region \citep{Rodney08} lying on the far side of the Aquila Rift, obscured by foreground extinction and by its own associated molecular gas and dust, in which it is still partly embedded. W40 is adjacent to the Serpens South molecular cloud \citep{Eiroa08,Shimoikura20}, which harbours several sites of ongoing  star formation, and appears to be a less evolved part of the same complex. The overall shape of W40 is best seen in infrared images beyond 2~$\mu$m (Fig.~\ref{W40_Spitzer}), which reveal a bipolar cavity crossed by a band of bright nebulosity that is also traced by molecular emission and is most clearly seen in the far-infrared images obtained by Herschel \citep{Bontemps10,Mallick13}. The pattern of illumination of the inner walls of the cavity clearly points to an ionization source located behind the band that separates both lobes. The structure is reminiscent of other bipolar HII regions, such as the more distant S106 \citep[][and references therein]{Schneider18}.  

Strong extinction in the direction of W40 and Serpens South prevented the identification of large numbers of young stellar objects in the region until relatively recent infrared surveys \citep{Gutermuth08,Povich13,Dunham15}. However, the existence of a compact group of infrared sources near the center of W40 was already reported by \citet{Zeilik78} and investigated in more detail by \citet{Smith85}, who made a first estimation of its contribution to the ionization of the nebula and discussed the evidence for abundant infrared excesses of several of its members. More recent near-infrared images (see for example Figure~2 of \citet{Kuhn10}) show a distinct compact cluster of highly reddened stars with an approximate diameter of $6'$. \citet{Shuping12} have confirmed the presence of at least a late O and several early B stars in the cluster, and have presented spectral classifications of some of its brightest members. Deep Chandra X-ray observations \citep{Kuhn10} have revealed over 200 young stellar objects in an area of $17' \times 17'$ centered on the cluster. Combining observations in the near-, mid- and far-infrared respectively obtained with the UK InfraRed Telescope (UKIRT) and the Spitzer and Herschel Space Observatories, \citet{Mallick13} confirm the existence of abundant YSOs all over the region, including some possible starless and pre-stellar cores distributed along the central lane separating the lobes of the nebula, first identified by \citet{Bontemps10} in Herschel observations. 

Obtaining a reliable census and characterization of the stars that contribute to the dissociation and ionization of the molecular gas in W40 is necessary to properly understand the physical properties of the HII region and its associated photodissociation region. W40 is a target of FEEDBACK \citep{Schneider20}, a Legacy Program being carried out with the airborne Stratospheric Observatory For Infrared Astronomy (SOFIA) to study stellar feedback in regions of massive star formation. FEEDBACK focuses on observations of the far-infrared cooling lines of ionized carbon at 158~$\mu$m and neutral oxygen at 63~$\mu$m in order to assess the radiative and mechanical energy input of massive stars into the interstellar medium.
  
In this paper we use the excellent astrometry delivered by the Gaia early Data Release 3 (eDR3) \citep{Gaiaedr321} to examine the membership of the W40 population,  accurately determine its distance, and confirm its extension beyond the boundaries of the central cluster. Many of the brightest members of the W40 stellar population are sufficiently bright at red-visible wavelengths to permit their spectral classification, which we present here.

\begin{figure}[ht]
\begin{center}
\hspace{-0.5cm}
\includegraphics [width=8.5cm, angle={0}]{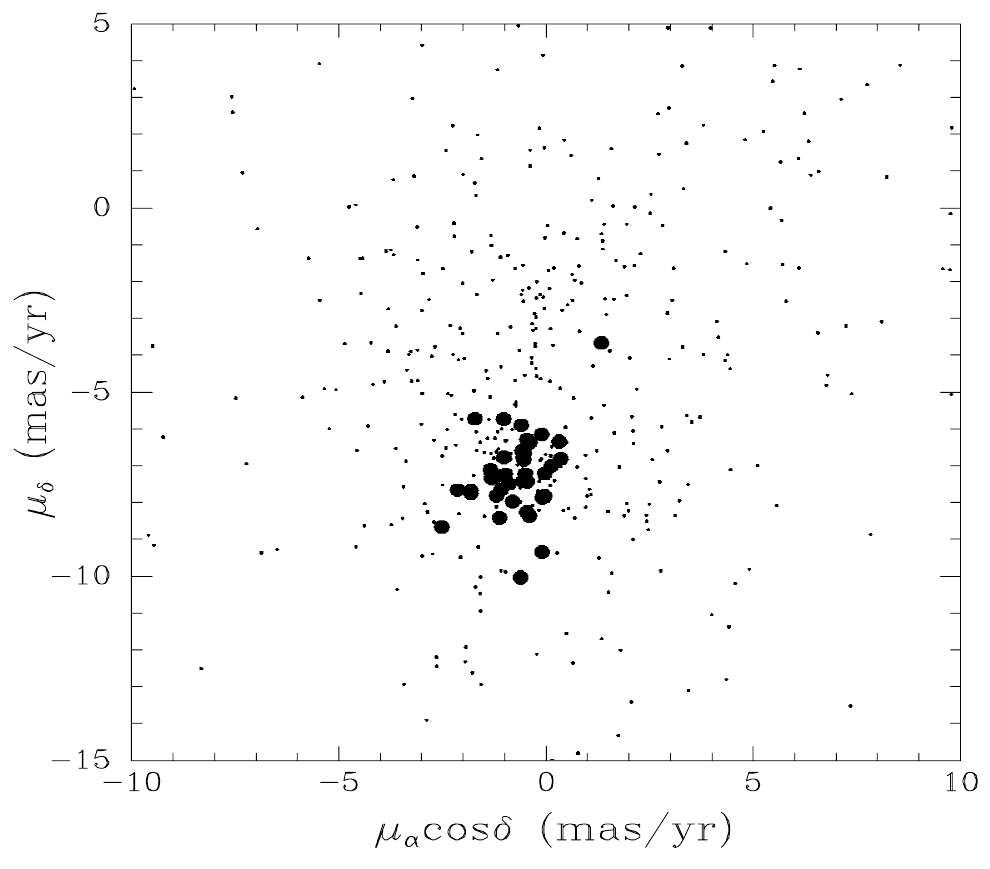}
\caption []{Proper motions of the stars within an area of $40' \times 40'$ centered on the coordinates $\alpha(2000) = 18^h31^m24^s$, $\delta(2000) = -02^\circ05'30''$ with Gaia eDR3 parallaxes between $1.0$~mas and $3.0$~mas. Stars within a radius of $2'$ from the central position are noted with filled circles.}
\label{plot_pm}
\end{center}
\end{figure}

\begin{figure}[ht]
\begin{center}
\hspace{-0.5cm}
\includegraphics [width=8.5cm, angle={0}]{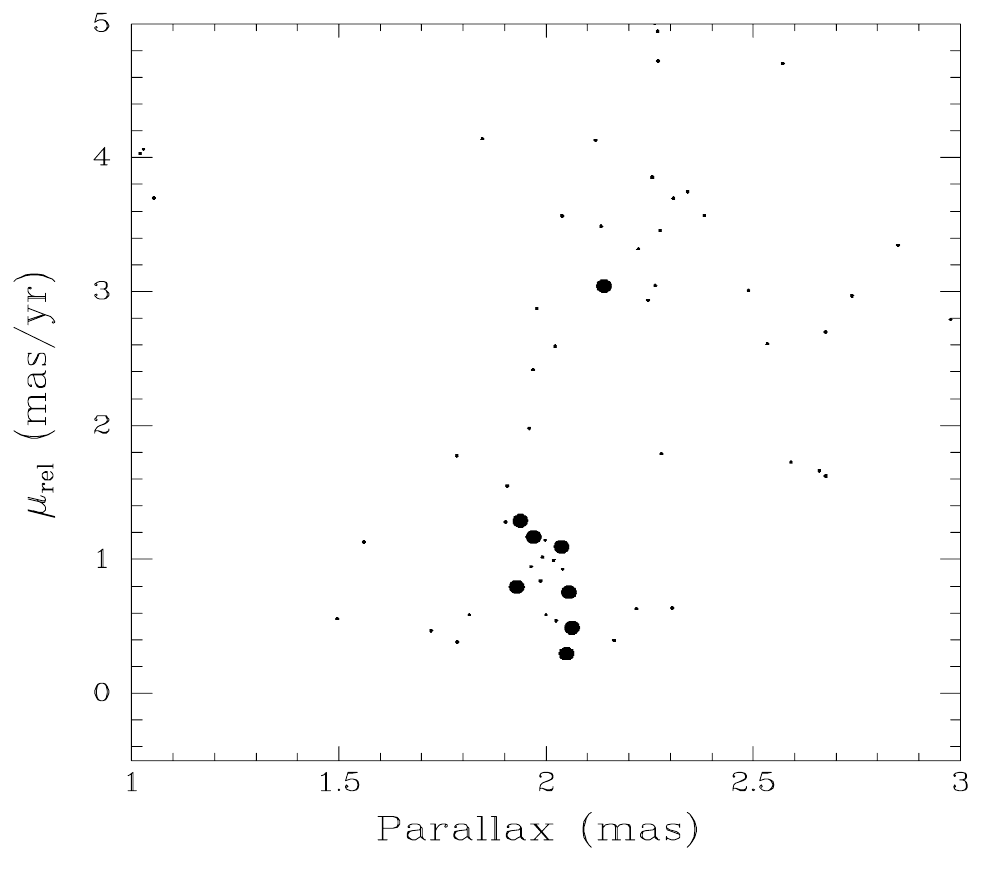}
\caption []{Residual proper motions with respect to $(\mu_\alpha \cos \delta, \mu_\delta) = (-0.8~{\rm mas~yr^{-1}}, -7.0~{\rm mas~yr^{-1}})$. Only stars with Gaia eDR3 parallax errors $\sigma(\pi) < 0.2$~mas and renormalized unit weight error (RUWE) below $1.4$ are plotted. Stars within a radius of $2'$ from the central position are noted with filled circles.}
\label{plot_parpm}
\end{center}
\end{figure}

\section{Astrometric selection of members\label{astrometry}}

\subsection{The distance to W40 \label{distance}}
 
A search of the Gaia eDR3 catalog returns 533 stars in a $40' \times 40'$ area around the approximate coordinates of the W40 central cluster, $\alpha(2000) = 18:31:24$, $\delta(2000) = -02:05:30$, with trigonometric parallaxes between $1.0$~mas and $3.0$~mas, without any constraints on the quality of the parallax measurements. This sample contains several highly reddened stars within the radius of the central cluster, as well as many candidate YSOs identified by \citet{Mallick13} and X-ray sources found by \citet{Kuhn10}.

Despite the generous limits in parallax of that selection, the kinematic signature of W40 is clearly visible in a proper motion diagram (Fig.~\ref{plot_pm}), especially when stars within only $2'$ of the central position are considered. From the strong concentration defined by the stars within the boundaries of the cluster we adopt $(\mu_\alpha \cos \delta, \mu_\delta) = (-0.8~{\rm mas~yr^{-1}}, -7.0~{\rm mas~yr^{-1}})$ as the average proper motion vector of the cluster. The concentration is less prominent but still noticeable among the stars outside the cluster boundaries. The proper motion clearly separates the W40 population from the Serpens star forming region identified by \citet{Dunham15}, whose proper motions are strongly concentrated around $(\mu_\alpha \cos \delta, \mu_\delta) = (+2.5~{\rm mas~yr^{-1}}, -8.0~{\rm mas~yr^{-1}})$ \citep{Herczeg19}.

To search for the equivalent signature in parallax, we have selected only stars with a parallar error $\sigma(\pi) < 0.2$~mas and a good astrometric solution, with a renormalized unit weight error (RUWE) below $1.4$ \citep{Lindegren18}, and have computed their residual proper motion with respect to the bulk proper motion of the cluster as $\mu_{\rm rel} = \sqrt{(\mu_\alpha \cos \delta + 0.8)^2 + (\mu_\delta + 7.0)^2}$. Figure~\ref{plot_parpm} shows the result, again with the stars within $2'$ from the cluster center highlighted. The presence of the W40 population is again obvious, with a peak density near 2~mas where most cluster members, and also many stars outside the cluster boundaries, are found. Using the sample of stars with $\mu_{\rm rel} < 2$~mas~yr$^{-1}$, $1.8 < \pi{(\rm mas)} < 2.2$, $\sigma(\pi) < 0.2$~mas, ${\rm RUWE} < 1.4$ within the $40' \times 40'$ area around the adopted cluster center yields a weighted average parallax $\pi = (1.993 \pm 0.014)$~mas, which translates to a distance $D = (502 \pm 4)$~pc. The 20 stars contributing to this sample are presented in Table~\ref{golden}, where the Gaia blue ($G_{BP}$) and red ($G_{RP}$) magnitudes are also given.

Past distance determinations to W40 have been subjected to large uncertainties \citep[see][for a historical review]{Rodney08}, ranging from less than 200~pc to as much as 800~pc. A much more accurate value of $(436 \pm 9)$~pc was derived by \citet{Ortiz17} based on VLBA measurements. This value was confirmed by \citet{Ortiz18}, who obtained $433^{+45}_{-38}$~pc based on Gaia DR2 data. It is therefore somewhat surprising that the value that we find differs by such a large amount, given the quality of the data used by those authors. We note that their selection of candidate W40 members is based on a two-step procedure, by which candidate YSOs in the region are first used to derive their mean proper motion, and those with individual proper motions deviating less than $3\sigma$ from the mean are then used to derive the mean parallax, with further filtering based on the reliability of their astrometry. Besides our use of the higher quality astrometry of Gaia eDR3, our procedure as outlined above uses instead the central cluster members as the primary reference for the determination of the mean proper motion, which in this way becomes less affected by possible contamination by foreground or background objects. Furthermore, we note that out of the five cluster members observed with VLBA by \citet{Ortiz17}, four (KGF 36, 97, 122, and 133) have unreliable parallax solutions as flagged by high values of the RUWE (4.98, 3.92, 3.70, and 1.44 respectively) and the fifth one, KGF~138, has a Gaia eDR3 parallax of $(1.9697 \pm 0.0778)$~mas, fully consistent with the value that we find for the overall population. The combination of the better data provided by Gaia eDR3, our selection criteria of W40 members somewhat less prone to possible contamination by non-members, and the problematic astrometry of most of the sources used in the VLBA determination may account for the differences with the work of \citet{Ortiz18}. We believe that the distance that we find, $D = (502 \pm 4)$~pc, can be adopted with higher confidence. We note that the uncertainty in the distance is similar to the projected size of W40, which is also $\sim 4$~pc across.

\begin{table*}
\caption{{W40 members with $\mu_{\rm rel} < 2$~mas~yr$^{-1}$, $1.8 < \pi{(\rm mas)} < 2.2$, $\sigma(\pi) < 0.2$~mas, ${\rm RUWE} < 1.4$}\label{golden}}
\begin{tabular}{ccccccccc}
\hline\hline
\noalign{\smallskip}
Star & $\pi$ & $\mu_\alpha \cos \delta$ & $\mu_\delta$ & RUWE & $G_{BP}$ & $G_{RP}$ & SVKY$^1$ & KGF$^2$ \\
             & (mas) & (mas~yr$^{1}$) & (mas~yr$^{1}$) & & & & nr. & nr. \\
\noalign{\smallskip}\hline\noalign{\smallskip}
J183105.38-020424.8 & $2.1640 \pm 0.1333$ & $-0.485 \pm 0.120$ & $-7.240 \pm 0.108$ & $1.070$ & $19.250 \pm 0.056$ & $14.234 \pm 0.004$ &    &     \\
J183122.58-020531.6 & $2.0491 \pm 0.0417$ & $-0.984 \pm 0.043$ & $-7.232 \pm 0.039$ & $1.310$ & $16.246 \pm 0.004$ & $12.500 \pm 0.004$ & 2B   &  87 \\
J183123.12-020521.0 & $2.0551 \pm 0.1955$ & $-0.387 \pm 0.190$ & $-6.368 \pm 0.184$ & $0.965$ & $20.999 \pm 0.165$ & $16.344 \pm 0.012$ &    &  93 \\
J183123.96-020410.8 & $2.0373 \pm 0.0443$ & $-0.106 \pm 0.047$ & $-6.154 \pm 0.044$ & $1.293$ & $15.438 \pm 0.004$ & $11.697 \pm 0.004$ & 3A   &     \\
J183124.82-022008.4 & $1.9641 \pm 0.0285$ & $+0.112 \pm 0.029$ & $-6.745 \pm 0.023$ & $1.177$ & $14.074 \pm 0.004$ & $11.326 \pm 0.004$ &    &     \\ 
J183124.93-020550.3 & $2.0626 \pm 0.1739$ & $-0.555 \pm 0.180$ & $-7.426 \pm 0.164$ & $1.137$ & $20.861 \pm 0.187$ & $16.266 \pm 0.013$ &    & 108 \\
J183125.43-020507.1 & $1.9292 \pm 0.1584$ & $-0.033 \pm 0.169$ & $-7.209 \pm 0.160$ & $1.135$ & $20.414 \pm 0.122$ & $16.042 \pm 0.029$ &    & 116 \\
J183127.66-020509.7 & $1.9697 \pm 0.0778$ & $+0.353 \pm 0.086$ & $-6.811 \pm 0.073$ & $1.303$ & $18.771 \pm 0.021$ & $14.168 \pm 0.004$ &    & 138 \\
J183128.00-020517.2 & $1.9378 \pm 0.0838$ & $-1.026 \pm 0.093$ & $-5.732 \pm 0.080$ & $1.193$ & $19.223 \pm 0.035$ & $14.618 \pm 0.005$ &    & 144 \\
J183136.58-015732.1 & $1.8146 \pm 0.1695$ & $-0.910 \pm 0.187$ & $-7.578 \pm 0.166$ & $1.076$ & $20.607 \pm 0.209$ & $16.429 \pm 0.007$ &    &     \\
J183139.90-021842.9 & $1.9907 \pm 0.0280$ & $-1.592 \pm 0.024$ & $-6.364 \pm 0.019$ & $1.168$ & $15.357 \pm 0.004$ & $12.890 \pm 0.004$ &    &     \\ 
J183140.76-020931.3 & $1.9977 \pm 0.0514$ & $+0.321 \pm 0.044$ & $-6.768 \pm 0.037$ & $1.144$ & $17.429 \pm 0.009$ & $14.265 \pm 0.004$ &    &     \\
J183144.16-021618.2 & $1.9866 \pm 0.1035$ & $-0.015 \pm 0.086$ & $-6.701 \pm 0.075$ & $1.009$ & $19.660 \pm 0.077$ & $15.389 \pm 0.006$ &    &     \\
J183149.20-021152.8 & $2.0182 \pm 0.0334$ & $-1.171 \pm 0.032$ & $-6.078 \pm 0.027$ & $1.381$ & $15.758 \pm 0.004$ & $12.496 \pm 0.004$ &    & 221 \\
J183153.40-020959.9 & $2.0400 \pm 0.1083$ & $-0.195 \pm 0.105$ & $-7.701 \pm 0.095$ & $0.993$ & $18.451 \pm 0.024$ & $14.326 \pm 0.006$ &    &     \\
J183156.29-020550.0 & $1.9020 \pm 0.1892$ & $+0.321 \pm 0.192$ & $-6.380 \pm 0.173$ & $0.962$ & $20.421 \pm 0.232$ & $16.822 \pm 0.011$ &    & 224 \\
J183157.53-020855.8 & $2.0243 \pm 0.1331$ & $-0.450 \pm 0.127$ & $-7.411 \pm 0.110$ & $0.948$ & $19.949 \pm 0.108$ & $16.393 \pm 0.016$ &    &     \\
J183204.02-021353.4 & $1.9995 \pm 0.0343$ & $-0.382 \pm 0.036$ & $-7.413 \pm 0.031$ & $1.243$ & $15.612 \pm 0.005$ & $12.027 \pm 0.004$ &    &     \\
J183216.77-022316.7 & $1.9060 \pm 0.1616$ & $+0.749 \pm 0.137$ & $-7.033 \pm 0.121$ & $1.054$ & $19.999 \pm 0.096$ & $16.212 \pm 0.014$ &    &     \\
J183240.25-022250.6 & $1.9591 \pm 0.0360$ & $+0.994 \pm 0.037$ & $-7.830 \pm 0.034$ & $1.056$ & $16.947 \pm 0.012$ & $13.579 \pm 0.004$ &    &     \\
\hline                            
\end{tabular}        
\smallskip\\
Notes:
\smallskip\\
$^1$: SVKY: Source identifyer in \citet{Shuping12}.\\
$^2$: KGF: Source identifyer in \citet{Kuhn10}.\\           
\end{table*}                      

\begin{figure*}[ht]
\begin{center}
\hspace{-0.5cm}
\includegraphics [width=12cm, angle={0}]{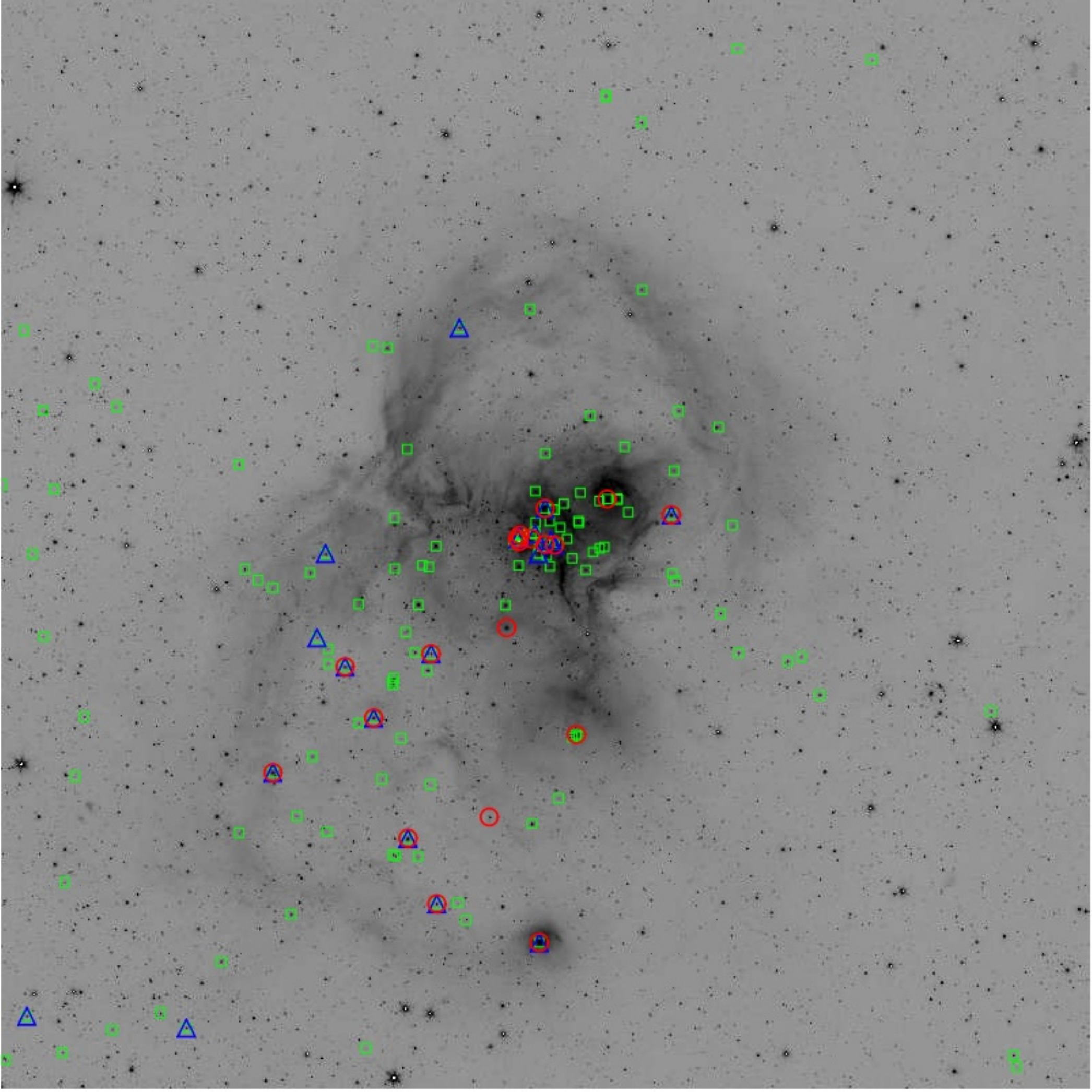}
\caption []{Astrometrically selected candidate members of the W40 population, superimposed on a Spitzer 4.5~$\mu$m image of the region. The size is $40'  \times 40'$ ($5.84 \times 5.84$~pc$^2$ projected at the distance of W40). Blue triangles are the high-confidence members having $\mu_{\rm rel} < 2$~mas~yr$^{-1}$,$1.8 < \pi{(\rm mas)} < 2.2$, $\sigma(\pi) < 0.2$~mas, ${\rm RUWE} < 1.4$ listed in Table~\ref{golden}. Green squares are the candidate members having $\mu_{\rm rel} < 0.2$~mas~yr$^{-1}$, $1.0< \pi{(\rm mas)} < 3.0$ with no restriction on the precision of the parallax or the quality of the astrometric solution listed in Table~\ref{faint}. The red circles indicate the stars for which we obtained visible spectroscopy as described in Section~\ref{observations}.}
\label{W40_YSOs}
\end{center}
\end{figure*}

The spatial distribution of high-confidence members listed in Table~\ref{golden} shows that the extended population reported in previous works is not an effect of superposition along the line of sight, but that such objects can be found all over the area covered by the nebula at present and even beyond. 

\subsection{Faint candidate members\label{faint}}

The precision with which Gaia can measure trigonometric parallaxes depends substantially on the magnitudes of the targets. This limits the sample of stars for which membership can be reliably established on the basis of both parallax and proper motion to the brightest or the least obscured members of W40. Nevertheless, proper motions still provide a viable membership selection criterion as long as separate populations with similar proper motions do no overlap along the line of sight, a caveat that must be kept in mind given the abundance of star forming regions at different distances in the Serpens-Aquila region \citep{Herczeg19}.

Figure~\ref{W40_YSOs} shows the positions of an additional 118 stars within the same limits of proper motion defined above, but now relaxing the limits of the parallax listed in Gaia eDR3 to the interval $1.0 < \pi ({\rm mas}) < 3.0$ and imposing no restriction on its uncertainty or the quality of the astrometric solution. These stars are generally fainter and therefore have higher parallax uncertainties than those of Table~\ref{golden}. While we may expect the list of high-confidence members in Table~\ref{golden} to be virtually free of contamination by non-members, this may not be entirely the case for the objects listed in Table~\ref{faint}. As shown by \citet{Herczeg19}, stars in the W40 / Serpens South region located at 300-500~pc based on Gaia~DR2 data show a large scatter in proper motion and may be a mixture resulting from different star formation episodes at different distances. A more detailed study of this population, able to elucidate its actual relationship with W40 and extending the high-confidence census of the latter toward fainter stars, should become possible with upcoming Gaia data releases.

\section{Spectroscopy\label{observations}}

The high-confidence list of W40 members presented in Table~\ref{golden} provides a sound basis for a more detailed study of the stars responsible for the ionization of the nebula. However, we may expect some actual intrinsically bright members to be excluded from that list due to heavy extinction, to particularly uncertain parallaxes due to their very red colors, or to problematic astrometric solutions resulting in a RUWE above the adopted threshold of $1.4$. The limit $\Delta \mu_{\rm rel} < 2$~mas~yr$^{-1}$ with respect to the bulk motion of the cluster that we use as a kinematic criterion for membership translates into a projected velocity of $4.8$~km~s$^{-1}$. This is similar to the internal velocity dispersion of much more massive star clusters and OB associations \citep{Rochau10,Melnik20}, but it may exclude stars that have obtained moderately high relative velocities due to three-body dynamical interactions interactions. 

We have selected for spectral classification the stars in Table~\ref{golden} with magnitude $G_{RP}< 16.0$ (with the only exception of J183240.25-022250.6, which is the most distant star from the center of the cluster), plus the 9 stars listed in Table~\ref{additional_Calar} which are likely members with problematic parallax measurements, or which have high quality parallaxes consistent with the distance discussed in Section~\ref{distance} but relative proper motion $\mu_{\rm rel}$ sightly above the 2~mas limit. The latter include J183144.27-021621.2 and 183127.81-020521.8, which are close companions to J183144.16-021618.2 and 183127.83-020523.7 respectively, and could be observed simultaneously with their primaries, although both have $G_{RP} > 16.0$.

\begin{table*}
\caption{Additional likely members selected for visible spectroscopy\label{additional_Calar}}
\begin{tabular}{ccccccccc}
\hline\hline
\noalign{\smallskip}
Star & $\pi$ & $\mu_\alpha \cos \delta$ & $\mu_\delta$ & RUWE & $G_{BP}$ & $G_{RP}$ & SVKY$^1$ & KGF$^2$ \\
             & (mas) & (mas~yr$^{1}$) & (mas~yr$^{1}$) & & & & nr. & nr. \\
\noalign{\smallskip}\hline\noalign{\smallskip}
183114.83-020350.0 & $1.5419 \pm 0.1177$ & $-0.177 \pm 0.115$ & $ -8.211 \pm 0.105$ & $4.980$ & $14.099 \pm  0.004$ & $10.799 \pm 0.005$ & 5   &  36 \\
183119.37-021229.5 & $1.9350 \pm 0.0953$ & $-0.691 \pm 0.074$ & $ -6.981 \pm 0.062$ & $3.153$ & $13.467 \pm  0.003$ & $10.601 \pm 0.004$ &     &     \\
183123.97-020529.5 & $1.7822 \pm 0.0953$ & $+0.317 \pm 0.096$ & $ -6.347 \pm 0.090$ & $2.358$ & $16.077 \pm  0.007$ & $12.039 \pm 0.007$ & 2A  &  99 \\
183126.02-020517.0 & $2.0323 \pm 0.3693$ & $-1.815 \pm 0.401$ & $ -7.683 \pm 0.342$ & $3.699$ & $19.846 \pm  0.057$ & $14.879 \pm 0.019$ & 1C  & 122 \\
183127.81-020521.8 & $2.1396 \pm 0.1025$ & $-0.616 \pm 0.112$ & $-10.035 \pm 0.096$ & $0.977$ & $                 $ & $                $ & 1A-N &     \\
183127.83-020523.7 & $2.4575 \pm 0.2611$ & $-0.035 \pm 0.285$ & $ -7.830 \pm 0.243$ & $6.319$ & $14.792 \pm  0.006$ & $10.839 \pm 0.007$ & 1A-S & 141 \\
183129.61-020833.8 & $2.0220 \pm 0.1171$ & $+0.264 \pm 0.112$ & $ -9.363 \pm 0.103$ & $1.387$ & $18.508 \pm  0.024$ & $13.982 \pm 0.004$ &     &     \\
183132.18-021531.2 & $1.9776 \pm 0.0835$ & $-1.092 \pm 0.066$ & $ -9.858 \pm 0.057$ & $0.991$ & $18.476 \pm  0.019$ & $15.498 \pm 0.007$ &     &     \\
183144.27-021621.2 & $3.5802 \pm 0.7898$ & $+0.325 \pm 0.680$ & $ -4.275 \pm 0.638$ & $2.474$ & $21.790 \pm  0.493$ & $16.935 \pm 0.051$ &     &     \\
\hline                            
\end{tabular}                    
\smallskip\\
Notes:
\smallskip\\
$^1$: SVKY: Source identifyer in \citet{Shuping12}.\\
$^2$: KGF: Source identifyer in \citet{Kuhn10}.\\           
\end{table*}  

The spectroscopic observations were carried out using CAFOS, the facility imager and spectrograph at the Calar Alto 2.2m telescope on the nights of 27 July, 28 July, and 8 August 2021. Two grism setups were used, having similar spectral resolutions ($\lambda / \Delta \lambda \simeq 700$ with the $1''2$-wide slit used) but different wavelength coverage. A blue grism covered the interval from 3800~\AA\ to 8700~\AA , whereas the red grism covered the interval from 6250~\AA\ to 10000~\AA . The use of the blue grism was intended to use spectral classification criteria at wavelengths shorter than 6000~\AA\ , which is possible only for a subset of our targets, and our classification is therefore largely based on the red spectral region common to all stars observed. Exposure times per star varied based on their red magnitude, and ranged from 900~s to 3600~s.  

\begin{figure*}[ht]
\begin{center}
\hspace{-0.5cm}
\includegraphics [width=14cm, angle={0}]{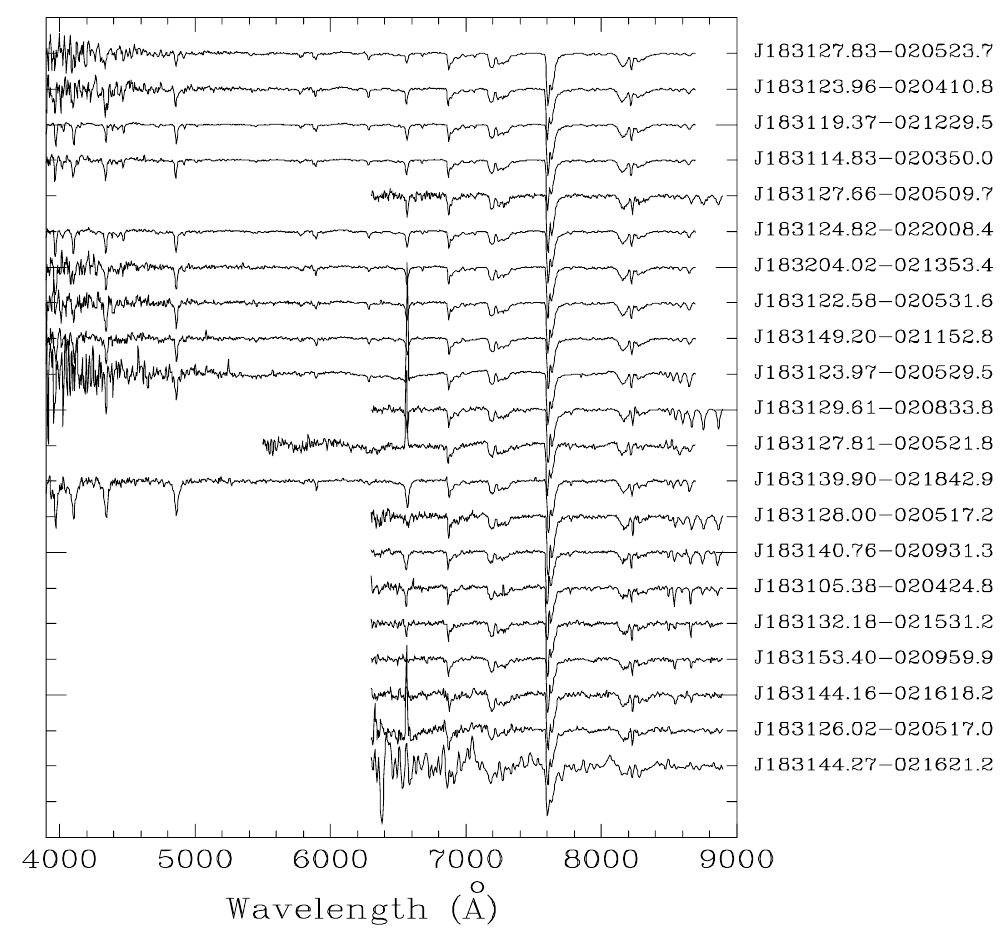}
\caption []{Continuum-normalized spectra of the stars listed in Table~\ref{physpar}, arranged from top to bottom from early to late types.}
\label{W40_spec}
\end{center}
\end{figure*}

\section{Results\label{results}}

\subsection{Spectral classification\label{classification}}

Despite the scarcity of absorption lines in low-resolution spectra of the red region of early-type stars, some useful diagnostics are present allowing a reasonably precise spectral classification. We use as a reference the atlas of \citet{Torres93}. In particular, the range covered by our spectra includes lines easily detectable when present at the resolution and typical signal-to-noise ratio of most of our spectra, such as H$\alpha$, the two HeI lines at 6678~\AA\ and 7064~\AA, the highest members of the Paschen series, the CaII infrared triplet, and the OI line at 7774~\AA\ . The varying behavior of those lines as a function of temperature, which is well illustrated by the atlas of \citet{Torres93}, combines to enable a spectral classification that we estimate to be precise to $\pm 2$ subtypes or better in the O to mid-F range, increasing up to 5 subtypes in the late-F to mid-K range. Although the precision of the classification should improve in the late-K and M range with the appearance of the TiO bands, the candidate late-type stars in our sample are too faint, and their spectra too noisy, to allow a classification with that degree of precision.

Figure~\ref{W40_spec} displays all our spectra arranged by spectral type, which is listed in Table~\ref{physpar}, normalized to a continuum fitted by a low-degree polynomial. Our spectral classifications expand and complement those of \citet{Shuping12}, who obtained near-infrared spectroscopy for eight members of the cluster. We have obtained red spectra of all of them with the exception of their source IRS-1B, which was too faint for our setup. We confirm their two Herbig Ae/Be stars, IRS~1A~North (=~J183127.81-02021.8), which we classify as A0:e, and IRS~2A (=~183123.7-020529.5), for which we obtain a B8e type. We classify IRS~1C as K-type, although the quality of the red spectrum prevents us from being more specific, hence our tentative classification as K:e. The late type is consistent with the presence of NaI, CaI, and CO in absorption in the K-band spectrum reported by \citet{Shuping12}, and the strong H$\alpha$ emission that we find is corresponded by emission lines in the Paschen and Brackett series found in their observations. \citet{Shuping12} give more precise classifications for another three B stars, which are in good agreement with ours: IRS~2B (=~J183122.58-020531.6), IRS~3A (=~J183123.96-020410.8), and IRS~5 (= J183114.83-020350.0), are respectively B5, B1, and B2 in their study, and B4, B3, and B1 in ours. Finally, both \citet{Shuping12} and our present study find IRS~1A-South (=~J183127.83-020523.7) to be the earliest-type member of the cluster. \citet{Shuping12} note that absence of HeII line at 2.1885~$\mu$m implies a spectral type later than O9, and our red spectrum shows a faint but clearly detected HeII line at 5411~\AA , confirming the late-O classification and consistent with the O9.5 type proposed by \citet{Shuping12}, which we adopt here. 

\begin{table*}
\caption{Intrinsic properties of stars with available spectroscopy\label{physpar}}
\begin{tabular}{lccccc}
\hline\hline
\noalign{\smallskip}
Star   & Sp. type & $\log T_{\rm eff}$ & $(G_{\rm BP} - G_{\rm RP})_0$ & 
$A_{G_{\rm RP}}$ & $M_{G_{\rm RP}}$\\    
\noalign{\smallskip}\hline\noalign{\smallskip}
J183105.38-020424.8     & A9       & 3.869 &  0.309 &  6.731 & -1.002 \\ 
J183114.83-020350.0$^1$ & B2       & 4.314 & -0.348 &  5.215 & -2.920 \\ 
J183119.37-021229.5     & B2       & 4.314 & -0.348 &  4.596 & -2.499 \\ 
J183122.58-020531.6$^1$ & B5       & 4.196 & -0.246 &  5.707 & -1.711 \\ 
J183123.96-020410.8$^1$ & B1       & 4.415 & -0.425 &  5.958 & -2.766 \\ 
J183123.97-020529.5$^1$ & B8e      & 4.090 & -0.156 &  5.997 & -2.462 \\  
J183124.82-022008.4     & B2       & 4.314 & -0.348 &  4.425 & -1.603 \\  
J183126.02-020517.0$^1$ & K:e      & 3.647 &  1.325 &  5.209 &  1.164 \\  
J183127.66-020509.7$^1$ & B2:      & 4.314 & -0.348 &  7.079 & -1.416 \\  
J183127.81-020521.8$^1$ & A0:e     &       &        &        &        \\
J183127.83-020523.7$^1$ & O9.5$^2$ & 4.501 & -0.487 &  6.350 & -4.016 \\  
J183128.00-020517.2$^1$ & B7       & 4.146 & -0.206 &  6.879 & -0.766 \\  
J183129.61-020833.8$^1$ & B8       & 4.090 & -0.156 &  6.693 & -1.215 \\
J183132.18-021531.2     & G5       & 3.753 &  0.820 &  3.085 &  3.907 \\
J183139.90-021842.9     & A5       & 3.908 &  0.145 &  3.319 &  1.066 \\  
J183140.76-020931.3     & A5       & 3.908 &  0.145 &  4.317 &  1.443 \\  
J183144.16-021618.2     & K7       & 3.613 &  1.547 &  3.893 &  2.991 \\  
J183144.27-021621.2     & M2:      & 3.551 &  2.290 &  3.667 &  4.763 \\
J183149.20-021152.8     & B5:      & 4.196 & -0.246 &  5.016 & -1.025 \\  
J183153.40-020959.9     & K2       & 3.708 &  1.016 &  4.445 &  1.375 \\  
J183204.02-021353.4     & B2       & 4.314 & -0.348 &  5.624 & -2.102 \\  
\hline                            
\end{tabular}
\smallskip\\
Notes:
\smallskip\\
$^1$: Member of the central cluster. \\
$^2$: Classification adopted from \citet{Shuping12}.\\
\end{table*} 

\subsection{Intrinsic properties\label{intrinsic}}

We have used the compilation of properties of main-sequence stars of \citet{Pecaut13} to link the spectral type to the effective temperature
$T_{\rm eff}$. We then use the MESA isochrone for 1~Myr 
\citep{Paxton11,Paxton13,Paxton15,Choi16,Dotter16} to estimate the 
the intrinsic $(G_{\rm BP} - G_{\rm RP})_0$ color from $T_{\rm eff}$ (see Table~\ref{physpar}). The $(G_{\rm BP} - G_{\rm RP})_0$ vs. $T_{\rm eff}$ is very weakly dependent on the adopted age, and the uncertainty resulting from our choice of the isochrone is fully absorbed within the uncertainty in the spectral classification. We estimate the extinction in $G_{\rm RP}$ using the extinction relations among different photometric bands derived by 
\citet{Wang19}:

\begin{equation}
A_{G_{\rm RP}} = 1.43 [(G_{\rm BP} - G_{\rm RP}) - (G_{\rm BP} - G_{\rm RP})_0]
\label{vi_transform}
\end{equation}

\noindent with $A_V = 1.676 A_{G_{\rm RP}}$. Using the distance modulus $DM = 8.504$ corresponding to the distance determined in Section~\ref{astrometry}, we compute the absolute magnitude $M_{G_{\rm RP}} = G_{\rm RP} - A_{G_{\rm RP}} - DM$. The derived values of the extinction and the absolute magnitude for each star are listed in Table~\ref{physpar}.

\begin{figure}[ht]
\begin{center}
\hspace{-0.5cm}
\includegraphics [width=8.5cm, angle={0}]{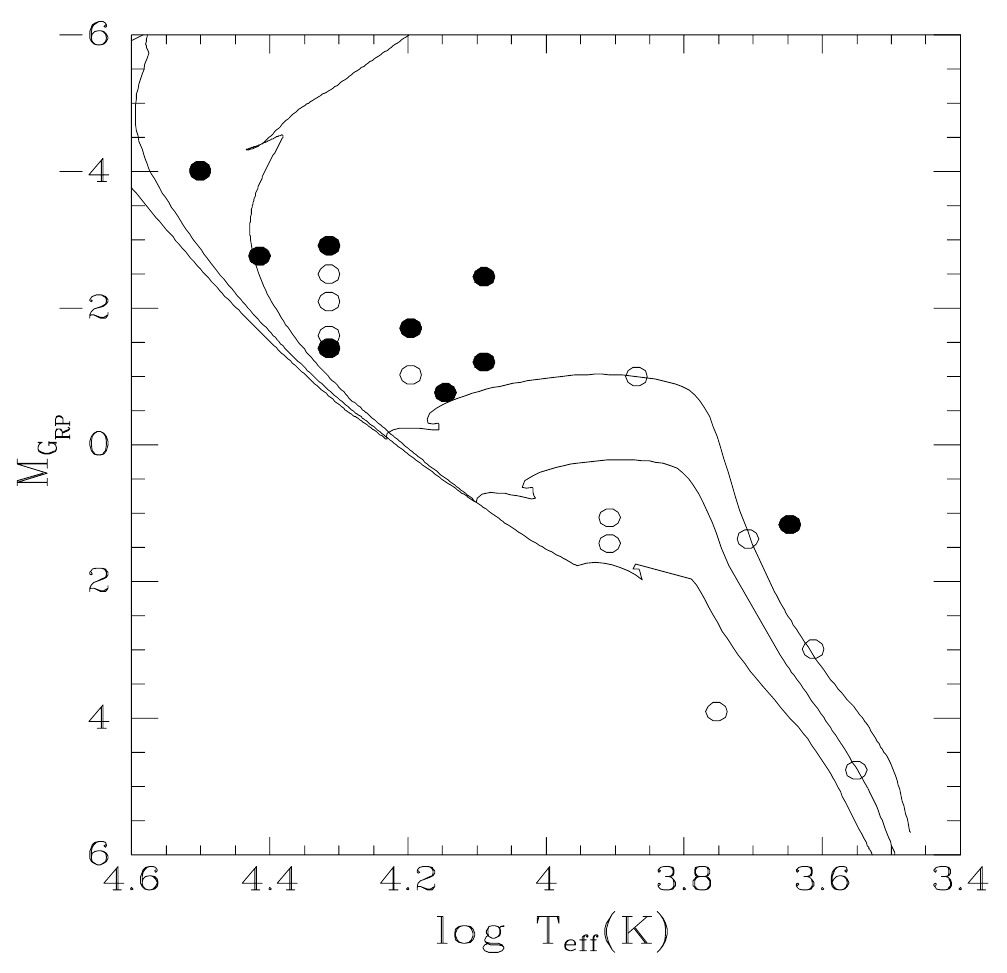}
\caption []{Temperature-magnitude diagram of all the stars in Table~\ref{physpar} (with the exception of J183127.81-020521.8, whose spectra type is poorly determined and no $G_{\rm BP}$ and $G_{\rm RP}$ photometry is available). The MESA isochrones for 1, 3 and 10~Myr are plotted as well. Full circles represent cluster members.} 
\label{hr}
\end{center}
\end{figure}

Figure~\ref{hr} shows the $\log T_{\rm eff}$ - $M_{G_{\rm RP}}$ diagram of the stars listed in Table~\ref{physpar}, with the exception of the Herbig Ae star J183127.81-020521.8, whose spectral type is poorly constrained and for which there are no magnitude measurements in the Gaia eDR3 catalog. Members of the central cluster and of the extended population are identified with different symbols. The location of the stars with respect to the MESA isochrones for 1, 3, and 10~Myr indicates a very young population, with only the most massive stars being near the main sequence. The location of the two earliest-type stars above it may furthermore be mainly due to similar-mass binarity, as it is frequently found among massive stars \citep{Sana11}. The position in the temperature-magnitude diagram of the cluster stars with masses below 4-5~M$_\odot$ (corresponding to mid-B types and later) suggests that the cluster is very young, almost certainly below 1~Myr, whereas the extended population may cover a wider range of ages. The latter conclusion is only tentative for the time being, as it is based on a small number of stars for which precise spectral types are particularly imprecise and therefore also are their derived temperatures and luminosities.

\begin{figure}[ht]
\begin{center}
\hspace{-0.5cm}
\includegraphics [width=8.5cm, angle={0}]{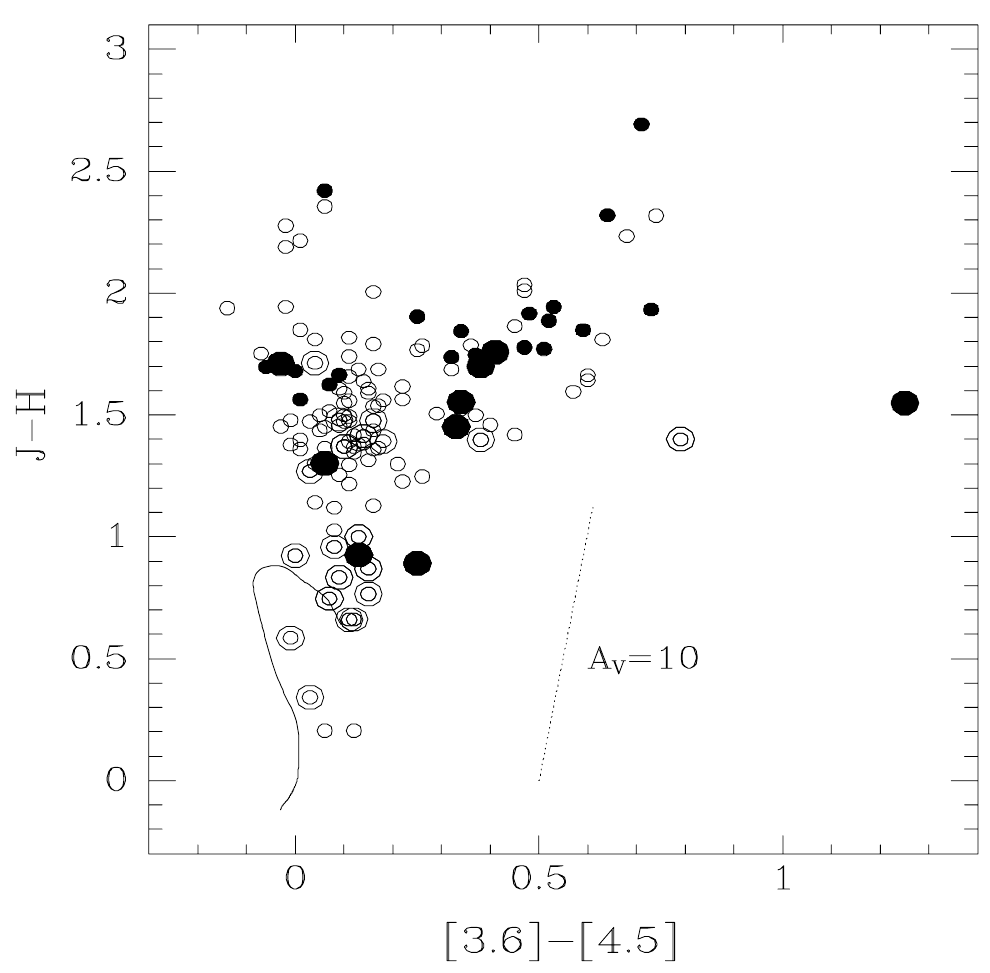}
\caption []{$(J-H, [3.6]-4.5])$ color-color diagram showing the locations of both high-confidence members of the W40 region (large symbols) and candidate members with Gaia eDR3 parallaxes between 1 and 3~mas (small symbols). Filled circles correspond to stars projected within the boundaries of the central cluster. The solid line is the predicted locus of the colors of unreddened stars with effective temperatures between $30,000$~K and $2,500$~K according to the MESA 1~Myr isochrone (which is virtually undistinguishable from the locus corresponding to other isochrones in the same temperature range). The dotted line is the reddening vector, with a length corresponding to $A_V = 10$~mag.}
\label{colcol}
\end{center}
\end{figure}

\subsection{Infrared excesses\label{ir_excess}}

We have used the 2MASS $JHK_S$ photometry together with the Spitzer IRAC photometry at $3.6$~$\mu$m and $4.5$~$\mu$m of \citet{Mallick13} to assess the presence of circumstellar material around the members of the population as an indicator of extreme youth. Figure~\ref{colcol} shows a color-color diagram where the $[3.6]-[4.5]$ index is mainly sensitive to the presence of excess continuum emission due to warm dust, whereas $J-H$ is mainly sensitive to the photospheric emission of the star and the foreground extinction. The locus occupied by unreddened stellar photospheres with temperatures ranging from $2,500$~K to $30,000$~K, from the synthetic MESA~1~Myr isochrone (see Section \ref{intrinsic}) is indicated, together with the reddening vector corresponding to an extinction $A_V = 10$~mag. Figure~\ref{colcol} displays the location of the sources in Tables~\ref{golden} and \ref{additional_Calar} (the high-confidence sample), as well as the candidate members with $\mu_{\rm rel} < 2$~mas and parallaxes $1.0 < \pi ({\rm mas}) < 3.0$ (see Section~\ref{faint}), with the stars projected on the central cluster noted with filled symbols.

We have established an approximate separation between stars without and with infrared excess based on their position left or right of the dividing line defined by a reddening vector having its origin at the position $([3.6]-[4.5], J-H) = (0.09, 0.64)$, corresponding to the colors of an unreddened photosphere at a temperature of $2,500$~K. Overall, most of the stars (57\%) plotted in Figure~\ref{colcol} lie in the no-excess region, but excesses are much more frequent among the stars projected within the cluster boundaries, where they reach 72\%, than among those belonging to the extended population, where they amount to 35\%. The fractions are similar when only the high-confidence members are considered: 10 out of 13 cluster members (77\%) display infrared excess, whereas only 12 out of the 28 stars (43\%) outside the cluster appear in that region of the color-color diagram.  

\section{Discussion \label{discussion}}

The classification of the members of the W40 cluster accessible to visible spectroscopy presented in Section~\ref{W40_spec} confirms J183127.83-020523.7 as its only O-type star, and therefore the main contributor to the ionizing radiation and the stellar wind that power the nebula and photodissociate its surrounding molecular cloud, with the caveat that it may be a similar-mass binary as hinted by its position in the temperature-absolute magnitude diagram. Besides J183127.83-020523.7 we only identify one B1 star, and possibly two B2 stars, as members of the cluster, while the remaining members for which we obtained spectra are of later types. The possible presence of significantly more obscured massive cluster members cannot be entirely ruled out, but we deem it unlikely. Nevertheless this point deserves further exploration through high signal-to-noise spectroscopy able to provide accurate classifications in the infrared \citep{Hanson96} of the most obscured cluster members. 

\begin{figure}[ht]
\begin{center}
\hspace{-0.5cm}
\includegraphics [width=8.5cm, angle={0}]{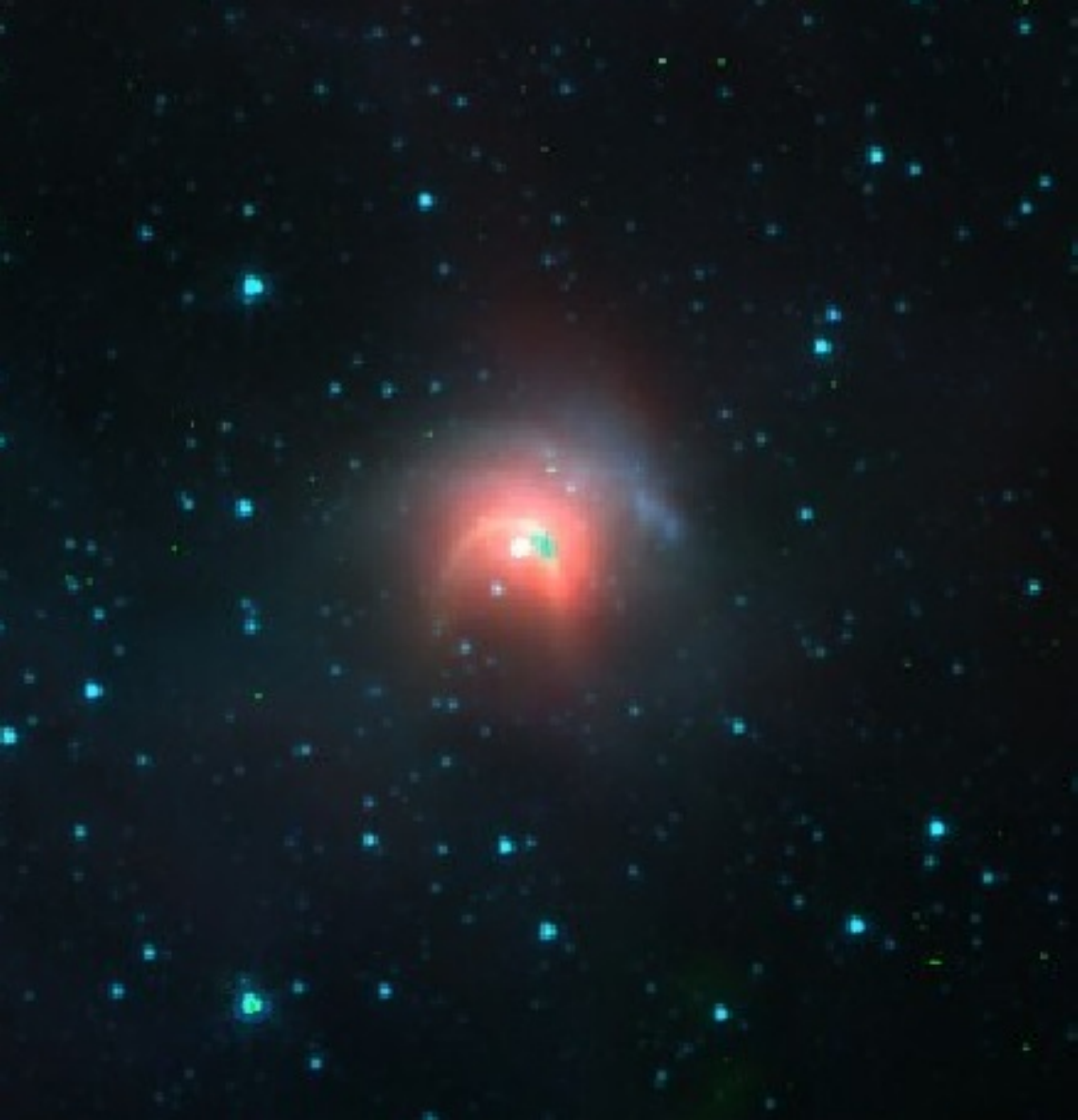}
\caption []{Surroundings of the B2 star J183124.82-022008.4, showing the arc-shaped structure that points approximately to the direction of the central cluster. Figure~\ref{W40_Spitzer} shows the position of this area with respect to the overall W40 region.  J183124.82-022008.4 is the bright star near the apsis of the arc. The arc is likely a bowshock produced where the ram pressure of the wind of the star balances the ram pressure of the expanding shell. Blue and green correspond respectively to the Spitzer IRAC 3.6~$\mu$m and 4.5~$\mu$m bands, and red to the MIPS 24~$\mu$m band.}
\label{bowshock}
\end{center}
\end{figure}

The morphology of bright rims and projected shadows at the inner edge of the cavity clearly points to the central cluster as the dominant source of illumination of the nebula. However, we find three B2 stars in the extended population which, although not being major contributors to the energetic input into the HII region, may cause important local effects and may have played an important role early in the history of the region. One of such local effects is clearly taking place at present, as shown by the arc-shaped structure surrounding the B2 star J18312.82-022008.4 (Fig.~\ref{bowshock}). The location of the star on the shell delimiting the extent of the bipolar nebula suggests a straightforward interpretation of the arc as a bow shock, formed as the gas in the expanding shell encounters the wind of the B star and is diverted around it, while the dust in the shell outside the arc is heated by the radiation of the star, giving rise to the local brightening of the nebula. 

The distribution of both the high-confidence members and of the candidates selected on proper motion is non-symmetric around the cluster. As Fig.~\ref{W40_YSOs} shows most of them are located in the southeastern quadrant, mostly within the contours of the nebula but also outside. The Chandra pointed observations of \citet{Kuhn10} do not cover a sufficiently wide area around the central cluster to clearly show this asymmetry, whereas the census of young stellar objects of \citet{Mallick13} is substantially incomplete given the low fraction of candidate members that display infrared excess.

The fractions of members with infrared excesses can be used as a crude approximation to estimate representative ages for the cluster and the extended population, due to the progressive dissipation of the inner disks \citep[see for instance][]{Kimura16,Yao18}. \citet{Richert18} have carried out an extensive analysis of infrared excess frequencies in the 1 to 8~$\mu$m range among X-ray selected samples in young clusters and aggregates sampling a wide range of ages $t$, and fitting the fraction $f$ of sources with infrared excess as an exponential decay $f = \exp (t / \tau)$. They find $\tau \simeq 1.9 - 2.8$~Myr, in broad agreement with the results of other authors (see \citet{Richert18} and references therein). This value depends nevertheless on whether or not the pre-main sequence isochrones used in their analysis include the effects of magnetic fields, rising to $\tau \simeq 5$~Myr for magnetic pre-main sequence models. Adopting $\tau = 2.8$~Myr, the fraction $f \simeq 0.72$ of stars with infrared excess that we obtain in the cluster suggests an age $t \simeq 0.9$~Myr, and the fraction $f \simeq 0.35$ found among the extended population corresponds to $t \simeq 3$~Myr. This is in qualitative agreement with the temperature-magnitude diagram (Fig.~\ref{hr}), taking into account the uncertainties involved in the location of stars in that diagram, the likely scatter of ages among the members of both populations, and the scatter in the disk fraction vs. age relationships. the youth of the cluster also agrees with the conclusions of \citet{Kuhn10}, who find that it has not yet reached dynamical relaxation although some degree of mass segregation has already taken place.   

It is interesting to note that while the older, extended population is mainly spread to the southwest of the W40 central cluster, the extremely young population composed of starless cores, protostars and Class I sources tends to concentrate toward the central lane and to the western part of the northern lobe of the nebula \citep{Mallick13,Konyves15}, as if indicating a progression of star formation with time from southeast to northwest. The presence of early-type stars among the older, extended population suggests that some degree of erosion of the interstellar medium in which the HII region is expanding may have taken place already before the birth of the W40 cluster. This offers an explanation to the greater extension of the southern lobe that is alternative to the one proposed by \citet{Mallick13}, who suggested the expansion of the northern lobe into a denser molecular medium as the cause for the difference in size between both.

We find no stars in the high-confidence sample with extinction below $A_V = 5$~mag, indicating that the entire region lies behind a thick layer of extinction. Using star counts in the direction of W40 / Serpens South combined with Gaia DR2 parallaxes \citet{Herczeg19} obtain a distance of $460 \pm 35$~pc for the clouds causing the extinction. The consistency of this estimate with the VLBI distance of W40 obtained by \citet{Ortiz17} lead \citet{Herczeg19} to propose the adoption of the latter, 436~pc, as the distance to the extinction layer, in which they consider W40 to be embedded. However, as discussed in Section~\ref{distance} we strongly favor a greater distance for W40, 502~pc, based on high-quality Gaia~eDR3 parallaxes of W40 cluster members. This places W40 behind the clouds responsible for the large foreground extinction, and possibly as a physically separated structure.

All the stars that we have observed in the W40 cluster are obscured by $A_V$ in the 8.7 to 11.7~mag range, indicating an additional contribution to the extinction around $A_V \simeq 4$~mag due to the dust lane that bisects the HII region. The physical association between the lane and the cluster is clear from the strong radiocontinuum emission peaking at the position of the cluster \citep{Mallick13}, indicating that the lane is being locally ionized by the cluster members. We note however that the obscuring material must have been already considerably disrupted and pierced, as the effects of the radiation of the massive stars of the cluster are noticeable in all the directions, including those aligned with the axis of the dust lane. We therefore propose that the additional contribution to the extinction in the direction of the cluster is caused by a remnant of the parental cloud lying along the line of sight, rather than by an undisrupted edge-on ring or torus obscuring the view of the cluster.

\section{Summary\label{summary}}

Our results can be summarized as follows:

\begin{itemize}

\item We present a redetermination of the distance to the W40 HII region based on high-quality Gaia eDR3 astrometry of members of its central cluster, which firmly establishes its distance as $(502 \pm 4)$~pc.

\item Astrometric criteria are used to identify stars in the region with proper motions and parallaxes very similar to those of the W40 central cluster. We identify 20 stars that we consider as high-confidence members of the region, 16 of which are outside the boundaries of the cluster, two outside the contour of the HII region, providing new evidence for an extended population.

\item We present spectroscopy in the red spectral region of 21 stars, 10 of them members of the cluster. We confirm or increase the accuracy of some previously published infrared-based classifications and provide new ones. We find new early-type stars, four of them having B2 spectral type and being located outside the cluster. One of these stars is interacting with the shell at the boundary of the HII region, producing a bow shock as its stellar wind collides with the expanding gas in the shell. Comparison with MESA pre-main sequence isochrones, the temperature-absolute magnitude diagram indicates an age younger than 1~Myr for the central cluster and hints to an older age for the extended population.

\item We identify 118 additional, generally fainter candidate members of the W40 region having proper motions very similar to those of the stars of the central cluster. Their membership to the region is less secure as the uncertainty in their parallaxes is higher,    but we note that their spatial distribution is similar to that of the high-confidence members of the region. 

\item Using published Spitzer photometry to identify stars with infrared excess due to circumstellar material, we find that approximately 2/3 of the cluster members lie in the region of the $(J-H, [3.6]-4.5])$ diagram indicative of the presence of excess, while the fraction drops to 1/3 for the extended population. Those fractions are very similar when the sample of high-confidence members and the sample of candidates with lower precision parallaxes are considered separately. Using the relationship between infrared excess fraction and stellar age found by other authors we estimate an age of $\sim 0.9$~Myr for the cluster, and $\sim 3$~Myr for the extended population.

\item All the members of the sample for which spectroscopy was obtained are obscured by more than 5 magnitudes of visual extinction. We attribute this obscuration to the presence of a foreground optically thick absorbing layer covering the entire region, which other works have estimated to lie at $(460 \pm 35)$~pc. 

\item All the W40 cluster members for which we have obtained spectra have extinction levels between $A_V \simeq 8.7$ and $A_V \simeq 11.7$~mag, indicating that the dust lane that bisects the HII region in mid- and far-infrared images 
adds approximately $\Delta A_V \sim 4$~mag of extinction. We argue however that the cluster is not fully embedded in this lane, as it does not prevent the ionizing radiation of the embedded stars from reaching virtually al the directions, including those along the axis of the lane.  

\end{itemize}

\begin{acknowledgements}

It is always a pleasure to thank the hospitality of the Calar Alto Observatory and the excellent support of its staff, on this occasion particularly of Jos\'e Ignacio Vico and Ana Guijarro. NS acknowledges support by the Agence National de Recherche (ANR/France) and the Deutsche Forschungsgemeinschaft (DFG/Germany) through the project “GENESIS” (ANR-16-CE92-0035-01/DFG1591/2-1). She also acknowledges support by the
BMWI via DLR, Project Number 50 OR 1916 (FEEDBACK) and the DFG project number SFB 956.
This work presents results from the European Space Agency (ESA) space mission Gaia. Gaia data are being processed by the Gaia Data Processing and Analysis Consortium (DPAC). Funding for the DPAC is provided by national institutions, in particular the institutions participating in the Gaia MultiLateral Agreement (MLA). This work is based in part on observations made with the Spitzer Space Telescope, which was operated by the Jet Propulsion Laboratory, California Institute of Technology under a contract with NASA. The Two Micron All Sky Survey (2MASS) is a joint project of the University of Massachusetts and the Infrared Processing and Analysis Center/California Institute of Technology, funded by NASA and the National Science Foundation. This research has made use of the SIMBAD database and the VizieR catalog service, both operated at CDS, Strasbourg, France.

\end{acknowledgements}

\bibliographystyle{aa} 
\bibliography{w40_cit}

\begin{thebibliography}{35}
\expandafter\ifx\csname natexlab\endcsname\relax\def\natexlab#1{#1}\fi

\bibitem[{{Bontemps} {et~al.}(2010){Bontemps}, {Andr{\'e}}, {K{\"o}nyves},
  {Men'shchikov}, {Schneider}, {Maury}, {Peretto}, {Arzoumanian}, {Attard},
  {Motte}, {Minier}, {Didelon}, {Saraceno}, {Abergel}, {Baluteau}, {Bernard},
  {Cambr{\'e}sy}, {Cox}, {di Francesco}, {di Giorgo}, {Griffin}, {Hargrave},
  {Huang}, {Kirk}, {Li}, {Martin}, {Mer{\'\i}n}, {Molinari}, {Olofsson},
  {Pezzuto}, {Prusti}, {Roussel}, {Russeil}, {Sauvage}, {Sibthorpe},
  {Spinoglio}, {Testi}, {Vavrek}, {Ward-Thompson}, {White}, {Wilson},
  {Woodcraft}, \& {Zavagno}}]{Bontemps10}
{Bontemps}, S., {Andr{\'e}}, P., {K{\"o}nyves}, V., {et~al.} 2010, \aap, 518,
  L85

\bibitem[{{Choi} {et~al.}(2016){Choi}, {Dotter}, {Conroy}, {Cantiello},
  {Paxton}, \& {Johnson}}]{Choi16}
{Choi}, J., {Dotter}, A., {Conroy}, C., {et~al.} 2016, \apj, 823, 102

\bibitem[{{Dotter}(2016)}]{Dotter16}
{Dotter}, A. 2016, \apjs, 222, 8

\bibitem[{{Dunham} {et~al.}(2015){Dunham}, {Allen}, {Evans},
  {Broekhoven-Fiene}, {Cieza}, {Di Francesco}, {Gutermuth}, {Harvey},
  {Hatchell}, {Heiderman}, {Huard}, {Johnstone}, {Kirk}, {Matthews}, {Miller},
  {Peterson}, \& {Young}}]{Dunham15}
{Dunham}, M.~M., {Allen}, L.~E., {Evans}, Neal~J., I., {et~al.} 2015, \apjs,
  220, 11

\bibitem[{{Eiroa} {et~al.}(2008){Eiroa}, {Djupvik}, \& {Casali}}]{Eiroa08}
{Eiroa}, C., {Djupvik}, A.~A., \& {Casali}, M.~M. 2008, in Handbook of Star
  Forming Regions, Volume II, ed. B.~{Reipurth}, Vol.~5, 693

\bibitem[{{Gaia Collaboration} {et~al.}(2021){Gaia Collaboration}, {Brown},
  {Vallenari}, {Prusti}, {de Bruijne}, {Babusiaux}, {Biermann}, {Creevey},
  {Evans}, {Eyer}, {Hutton}, {Jansen}, {Jordi}, {Klioner}, {Lammers},
  {Lindegren}, {Luri}, {Mignard}, {Panem}, {Pourbaix}, {Randich}, {Sartoretti},
  {Soubiran}, {Walton}, {Arenou}, {Bailer-Jones}, {Bastian}, {Cropper},
  {Drimmel}, {Katz}, {Lattanzi}, {van Leeuwen}, {Bakker}, {Cacciari},
  {Casta{\~n}eda}, {De Angeli}, {Ducourant}, {Fabricius}, {Fouesneau},
  {Fr{\'e}mat}, {Guerra}, {Guerrier}, {Guiraud}, {Jean-Antoine Piccolo},
  {Masana}, {Messineo}, {Mowlavi}, {Nicolas}, {Nienartowicz}, {Pailler},
  {Panuzzo}, {Riclet}, {Roux}, {Seabroke}, {Sordo}, {Tanga}, {Th{\'e}venin},
  {Gracia-Abril}, {Portell}, {Teyssier}, {Altmann}, {Andrae}, {Bellas-Velidis},
  {Benson}, {Berthier}, {Blomme}, {Brugaletta}, {Burgess}, {Busso}, {Carry},
  {Cellino}, {Cheek}, {Clementini}, {Damerdji}, {Davidson}, {Delchambre},
  {Dell'Oro}, {Fern{\'a}ndez-Hern{\'a}ndez}, {Galluccio}, {Garc{\'\i}a-Lario},
  {Garcia-Reinaldos}, {Gonz{\'a}lez-N{\'u}{\~n}ez}, {Gosset}, {Haigron},
  {Halbwachs}, {Hambly}, {Harrison}, {Hatzidimitriou}, {Heiter},
  {Hern{\'a}ndez}, {Hestroffer}, {Hodgkin}, {Holl}, {Jan{\ss}en}, {Jevardat de
  Fombelle}, {Jordan}, {Krone-Martins}, {Lanzafame}, {L{\"o}ffler}, {Lorca},
  {Manteiga}, {Marchal}, {Marrese}, {Moitinho}, {Mora}, {Muinonen}, {Osborne},
  {Pancino}, {Pauwels}, {Petit}, {Recio-Blanco}, {Richards}, {Riello},
  {Rimoldini}, {Robin}, {Roegiers}, {Rybizki}, {Sarro}, {Siopis}, {Smith},
  {Sozzetti}, {Ulla}, {Utrilla}, {van Leeuwen}, {van Reeven}, {Abbas}, {Abreu
  Aramburu}, {Accart}, {Aerts}, {Aguado}, {Ajaj}, {Altavilla}, {{\'A}lvarez},
  {{\'A}lvarez Cid-Fuentes}, {Alves}, {Anderson}, {Anglada Varela}, {Antoja},
  {Audard}, {Baines}, {Baker}, {Balaguer-N{\'u}{\~n}ez}, {Balbinot}, {Balog},
  {Barache}, {Barbato}, {Barros}, {Barstow}, {Bartolom{\'e}}, {Bassilana},
  {Bauchet}, {Baudesson-Stella}, {Becciani}, {Bellazzini}, {Bernet}, {Bertone},
  {Bianchi}, {Blanco-Cuaresma}, {Boch}, {Bombrun}, {Bossini}, {Bouquillon},
  {Bragaglia}, {Bramante}, {Breedt}, {Bressan}, {Brouillet}, {Bucciarelli},
  {Burlacu}, {Busonero}, {Butkevich}, {Buzzi}, {Caffau}, {Cancelliere},
  {C{\'a}novas}, {Cantat-Gaudin}, {Carballo}, {Carlucci}, {Carnerero},
  {Carrasco}, {Casamiquela}, {Castellani}, {Castro-Ginard}, {Castro Sampol},
  {Chaoul}, {Charlot}, {Chemin}, {Chiavassa}, {Cioni}, {Comoretto}, {Cooper},
  {Cornez}, {Cowell}, {Crifo}, {Crosta}, {Crowley}, {Dafonte}, {Dapergolas},
  {David}, {David}, {de Laverny}, {De Luise}, {De March}, {De Ridder}, {de
  Souza}, {de Teodoro}, {de Torres}, {del Peloso}, {del Pozo}, {Delbo},
  {Delgado}, {Delgado}, {Delisle}, {Di Matteo}, {Diakite}, {Diener},
  {Distefano}, {Dolding}, {Eappachen}, {Edvardsson}, {Enke}, {Esquej}, {Fabre},
  {Fabrizio}, {Faigler}, {Fedorets}, {Fernique}, {Fienga}, {Figueras},
  {Fouron}, {Fragkoudi}, {Fraile}, {Franke}, {Gai}, {Garabato},
  {Garcia-Gutierrez}, {Garc{\'\i}a-Torres}, {Garofalo}, {Gavras}, {Gerlach},
  {Geyer}, {Giacobbe}, {Gilmore}, {Girona}, {Giuffrida}, {Gomel}, {Gomez},
  {Gonzalez-Santamaria}, {Gonz{\'a}lez-Vidal}, {Granvik},
  {Guti{\'e}rrez-S{\'a}nchez}, {Guy}, {Hauser}, {Haywood}, {Helmi}, {Hidalgo},
  {Hilger}, {H{\l}adczuk}, {Hobbs}, {Holland}, {Huckle}, {Jasniewicz},
  {Jonker}, {Juaristi Campillo}, {Julbe}, {Karbevska}, {Kervella}, {Khanna},
  {Kochoska}, {Kontizas}, {Kordopatis}, {Korn}, {Kostrzewa-Rutkowska},
  {Kruszy{\'n}ska}, {Lambert}, {Lanza}, {Lasne}, {Le Campion}, {Le Fustec},
  {Lebreton}, {Lebzelter}, {Leccia}, {Leclerc}, {Lecoeur-Taibi}, {Liao},
  {Licata}, {Lindstr{\o}m}, {Lister}, {Livanou}, {Lobel}, {Madrero Pardo},
  {Managau}, {Mann}, {Marchant}, {Marconi}, {Marcos Santos}, {Marinoni},
  {Marocco}, {Marshall}, {Martin Polo}, {Mart{\'\i}n-Fleitas}, {Masip},
  {Massari}, {Mastrobuono-Battisti}, {Mazeh}, {McMillan}, {Messina},
  {Michalik}, {Millar}, {Mints}, {Molina}, {Molinaro}, {Moln{\'a}r},
  {Montegriffo}, {Mor}, {Morbidelli}, {Morel}, {Morris}, {Mulone}, {Munoz},
  {Muraveva}, {Murphy}, {Musella}, {Noval}, {Ord{\'e}novic}, {Orr{\`u}},
  {Osinde}, {Pagani}, {Pagano}, {Palaversa}, {Palicio}, {Panahi}, {Pawlak},
  {Pe{\~n}alosa Esteller}, {Penttil{\"a}}, {Piersimoni}, {Pineau}, {Plachy},
  {Plum}, {Poggio}, {Poretti}, {Poujoulet}, {Pr{\v{s}}a}, {Pulone}, {Racero},
  {Ragaini}, {Rainer}, {Raiteri}, {Rambaux}, {Ramos}, {Ramos-Lerate}, {Re
  Fiorentin}, {Regibo}, {Reyl{\'e}}, {Ripepi}, {Riva}, {Rixon}, {Robichon},
  {Robin}, {Roelens}, {Rohrbasser}, {Romero-G{\'o}mez}, {Rowell}, {Royer},
  {Rybicki}, {Sadowski}, {Sagrist{\`a} Sell{\'e}s}, {Sahlmann}, {Salgado},
  {Salguero}, {Samaras}, {Sanchez Gimenez}, {Sanna}, {Santove{\~n}a},
  {Sarasso}, {Schultheis}, {Sciacca}, {Segol}, {Segovia}, {S{\'e}gransan},
  {Semeux}, {Shahaf}, {Siddiqui}, {Siebert}, {Siltala}, {Slezak}, {Smart},
  {Solano}, {Solitro}, {Souami}, {Souchay}, {Spagna}, {Spoto}, {Steele},
  {Steidelm{\"u}ller}, {Stephenson}, {S{\"u}veges}, {Szabados}, {Szegedi-Elek},
  {Taris}, {Tauran}, {Taylor}, {Teixeira}, {Thuillot}, {Tonello}, {Torra},
  {Torra}, {Turon}, {Unger}, {Vaillant}, {van Dillen}, {Vanel}, {Vecchiato},
  {Viala}, {Vicente}, {Voutsinas}, {Weiler}, {Wevers}, {Wyrzykowski}, {Yoldas},
  {Yvard}, {Zhao}, {Zorec}, {Zucker}, {Zurbach}, \& {Zwitter}}]{Gaiaedr321}
{Gaia Collaboration}, {Brown}, A.~G.~A., {Vallenari}, A., {et~al.} 2021, \aap,
  649, A1

\bibitem[{{Gutermuth} {et~al.}(2008){Gutermuth}, {Bourke}, {Allen}, {Myers},
  {Megeath}, {Matthews}, {J{\o}rgensen}, {Di Francesco}, {Ward-Thompson},
  {Huard}, {Brooke}, {Dunham}, {Cieza}, {Harvey}, \& {Chapman}}]{Gutermuth08}
{Gutermuth}, R.~A., {Bourke}, T.~L., {Allen}, L.~E., {et~al.} 2008, \apjl, 673,
  L151

\bibitem[{{Hanson} {et~al.}(1996){Hanson}, {Conti}, \& {Rieke}}]{Hanson96}
{Hanson}, M.~M., {Conti}, P.~S., \& {Rieke}, M.~J. 1996, \apjs, 107, 281

\bibitem[{{Herczeg} {et~al.}(2019){Herczeg}, {Kuhn}, {Zhou}, {Hatchell},
  {Manara}, {Johnstone}, {Dunham}, {Bhardwaj}, {Jose}, \& {Yuan}}]{Herczeg19}
{Herczeg}, G.~J., {Kuhn}, M.~A., {Zhou}, X., {et~al.} 2019, \apj, 878, 111

\bibitem[{{Kimura} {et~al.}(2016){Kimura}, {Kunitomo}, \&
  {Takahashi}}]{Kimura16}
{Kimura}, S.~S., {Kunitomo}, M., \& {Takahashi}, S.~Z. 2016, \mnras, 461, 2257

\bibitem[{{K{\"o}nyves} {et~al.}(2015){K{\"o}nyves}, {Andr{\'e}},
  {Men'shchikov}, {Palmeirim}, {Arzoumanian}, {Schneider}, {Roy}, {Didelon},
  {Maury}, {Shimajiri}, {Di Francesco}, {Bontemps}, {Peretto}, {Benedettini},
  {Bernard}, {Elia}, {Griffin}, {Hill}, {Kirk}, {Ladjelate}, {Marsh}, {Martin},
  {Motte}, {Nguy{\^e}n Luong}, {Pezzuto}, {Roussel}, {Rygl}, {Sadavoy},
  {Schisano}, {Spinoglio}, {Ward-Thompson}, \& {White}}]{Konyves15}
{K{\"o}nyves}, V., {Andr{\'e}}, P., {Men'shchikov}, A., {et~al.} 2015, \aap,
  584, A91

\bibitem[{{Kuhn} {et~al.}(2010){Kuhn}, {Getman}, {Feigelson}, {Reipurth},
  {Rodney}, \& {Garmire}}]{Kuhn10}
{Kuhn}, M.~A., {Getman}, K.~V., {Feigelson}, E.~D., {et~al.} 2010, \apj, 725,
  2485

\bibitem[{{Lindegren} {et~al.}(2018){Lindegren}, {Hern{\'a}ndez}, {Bombrun},
  {Klioner}, {Bastian}, {Ramos-Lerate}, {de Torres}, {Steidelm{\"u}ller},
  {Stephenson}, {Hobbs}, {Lammers}, {Biermann}, {Geyer}, {Hilger}, {Michalik},
  {Stampa}, {McMillan}, {Casta{\~n}eda}, {Clotet}, {Comoretto}, {Davidson},
  {Fabricius}, {Gracia}, {Hambly}, {Hutton}, {Mora}, {Portell}, {van Leeuwen},
  {Abbas}, {Abreu}, {Altmann}, {Andrei}, {Anglada}, {Balaguer-N{\'u}{\~n}ez},
  {Barache}, {Becciani}, {Bertone}, {Bianchi}, {Bouquillon}, {Bourda},
  {Br{\"u}semeister}, {Bucciarelli}, {Busonero}, {Buzzi}, {Cancelliere},
  {Carlucci}, {Charlot}, {Cheek}, {Crosta}, {Crowley}, {de Bruijne}, {de
  Felice}, {Drimmel}, {Esquej}, {Fienga}, {Fraile}, {Gai}, {Garralda},
  {Gonz{\'a}lez-Vidal}, {Guerra}, {Hauser}, {Hofmann}, {Holl}, {Jordan},
  {Lattanzi}, {Lenhardt}, {Liao}, {Licata}, {Lister}, {L{\"o}ffler},
  {Marchant}, {Martin-Fleitas}, {Messineo}, {Mignard}, {Morbidelli}, {Poggio},
  {Riva}, {Rowell}, {Salguero}, {Sarasso}, {Sciacca}, {Siddiqui}, {Smart},
  {Spagna}, {Steele}, {Taris}, {Torra}, {van Elteren}, {van Reeven}, \&
  {Vecchiato}}]{Lindegren18}
{Lindegren}, L., {Hern{\'a}ndez}, J., {Bombrun}, A., {et~al.} 2018, \aap, 616,
  A2

\bibitem[{{Mallick} {et~al.}(2013){Mallick}, {Kumar}, {Ojha}, {Bachiller},
  {Samal}, \& {Pirogov}}]{Mallick13}
{Mallick}, K.~K., {Kumar}, M.~S.~N., {Ojha}, D.~K., {et~al.} 2013, \apj, 779,
  113

\bibitem[{{Melnik} \& {Dambis}(2020)}]{Melnik20}
{Melnik}, A.~M. \& {Dambis}, A.~K. 2020, \mnras, 493, 2339

\bibitem[{{Ortiz-Le{\'o}n} {et~al.}(2017){Ortiz-Le{\'o}n}, {Dzib}, {Kounkel},
  {Loinard}, {Mioduszewski}, {Rodr{\'\i}guez}, {Torres}, {Pech}, {Rivera},
  {Hartmann}, {Boden}, {Evans}, {Brice{\~n}o}, {Tobin}, \& {Galli}}]{Ortiz17}
{Ortiz-Le{\'o}n}, G.~N., {Dzib}, S.~A., {Kounkel}, M.~A., {et~al.} 2017, \apj,
  834, 143

\bibitem[{{Ortiz-Le{\'o}n} {et~al.}(2018){Ortiz-Le{\'o}n}, {Loinard}, {Dzib},
  {Kounkel}, {Galli}, {Tobin}, {Evans}, {Hartmann}, {Rodr{\'\i}guez},
  {Brice{\~n}o}, {Torres}, \& {Mioduszewski}}]{Ortiz18}
{Ortiz-Le{\'o}n}, G.~N., {Loinard}, L., {Dzib}, S.~A., {et~al.} 2018, \apjl,
  869, L33

\bibitem[{{Paxton} {et~al.}(2011){Paxton}, {Bildsten}, {Dotter}, {Herwig},
  {Lesaffre}, \& {Timmes}}]{Paxton11}
{Paxton}, B., {Bildsten}, L., {Dotter}, A., {et~al.} 2011, \apjs, 192, 3

\bibitem[{{Paxton} {et~al.}(2013){Paxton}, {Cantiello}, {Arras}, {Bildsten},
  {Brown}, {Dotter}, {Mankovich}, {Montgomery}, {Stello}, {Timmes}, \&
  {Townsend}}]{Paxton13}
{Paxton}, B., {Cantiello}, M., {Arras}, P., {et~al.} 2013, \apjs, 208, 4

\bibitem[{{Paxton} {et~al.}(2015){Paxton}, {Marchant}, {Schwab}, {Bauer},
  {Bildsten}, {Cantiello}, {Dessart}, {Farmer}, {Hu}, {Langer}, {Townsend},
  {Townsley}, \& {Timmes}}]{Paxton15}
{Paxton}, B., {Marchant}, P., {Schwab}, J., {et~al.} 2015, \apjs, 220, 15

\bibitem[{{Pecaut} \& {Mamajek}(2013)}]{Pecaut13}
{Pecaut}, M.~J. \& {Mamajek}, E.~E. 2013, \apjs, 208, 9

\bibitem[{{Povich} {et~al.}(2013){Povich}, {Kuhn}, {Getman}, {Busk},
  {Feigelson}, {Broos}, {Townsley}, {King}, \& {Naylor}}]{Povich13}
{Povich}, M.~S., {Kuhn}, M.~A., {Getman}, K.~V., {et~al.} 2013, \apjs, 209, 31

\bibitem[{{Richert} {et~al.}(2018){Richert}, {Getman}, {Feigelson}, {Kuhn},
  {Broos}, {Povich}, {Bate}, \& {Garmire}}]{Richert18}
{Richert}, A.~J.~W., {Getman}, K.~V., {Feigelson}, E.~D., {et~al.} 2018,
  \mnras, 477, 5191

\bibitem[{{Rochau} {et~al.}(2010){Rochau}, {Brandner}, {Stolte}, {Gennaro},
  {Gouliermis}, {Da Rio}, {Dzyurkevich}, \& {Henning}}]{Rochau10}
{Rochau}, B., {Brandner}, W., {Stolte}, A., {et~al.} 2010, \apjl, 716, L90

\bibitem[{{Rodney} \& {Reipurth}(2008)}]{Rodney08}
{Rodney}, S.~A. \& {Reipurth}, B. 2008, in Handbook of Star Forming Regions,
  Volume II, ed. B.~{Reipurth}, Vol.~5, 683

\bibitem[{{Sana} \& {Evans}(2011)}]{Sana11}
{Sana}, H. \& {Evans}, C.~J. 2011, in Active OB Stars: Structure, Evolution,
  Mass Loss, and Critical Limits, ed. C.~{Neiner}, G.~{Wade}, G.~{Meynet}, \&
  G.~{Peters}, Vol. 272, 474--485

\bibitem[{{Schneider} {et~al.}(2018){Schneider}, {R{\"o}llig}, {Simon},
  {Wiesemeyer}, {Gusdorf}, {Stutzki}, {G{\"u}sten}, {Bontemps}, {Comer{\'o}n},
  {Csengeri}, {Adams}, \& {Richter}}]{Schneider18}
{Schneider}, N., {R{\"o}llig}, M., {Simon}, R., {et~al.} 2018, \aap, 617, A45

\bibitem[{{Schneider} {et~al.}(2020){Schneider}, {Simon}, {Guevara},
  {Buchbender}, {Higgins}, {Okada}, {Stutzki}, {G{\"u}sten}, {Anderson},
  {Bally}, {Beuther}, {Bonne}, {Bontemps}, {Chambers}, {Csengeri}, {Graf},
  {Gusdorf}, {Jacobs}, {Justen}, {Kabanovic}, {Karim}, {Luisi}, {Menten},
  {Mertens}, {Mookerjea}, {Ossenkopf-Okada}, {Pabst}, {Pound}, {Richter},
  {Reyes}, {Ricken}, {R{\"o}llig}, {Russeil}, {S{\'a}nchez-Monge}, {Sandell},
  {Tiwari}, {Wiesemeyer}, {Wolfire}, {Wyrowski}, {Zavagno}, \&
  {Tielens}}]{Schneider20}
{Schneider}, N., {Simon}, R., {Guevara}, C., {et~al.} 2020, \pasp, 132, 104301

\bibitem[{{Shimoikura} {et~al.}(2020){Shimoikura}, {Dobashi}, {Hatano}, \&
  {Nakamura}}]{Shimoikura20}
{Shimoikura}, T., {Dobashi}, K., {Hatano}, Y., \& {Nakamura}, F. 2020, \apj,
  895, 137

\bibitem[{{Shuping} {et~al.}(2012){Shuping}, {Vacca}, {Kassis}, \&
  {Yu}}]{Shuping12}
{Shuping}, R.~Y., {Vacca}, W.~D., {Kassis}, M., \& {Yu}, K.~C. 2012, \aj, 144,
  116

\bibitem[{{Smith} {et~al.}(1985){Smith}, {Bentley}, {Castelaz}, {Gehrz},
  {Grasdalen}, \& {Hackwell}}]{Smith85}
{Smith}, J., {Bentley}, A., {Castelaz}, M., {et~al.} 1985, \apj, 291, 571

\bibitem[{{Torres-Dodgen} \& {Weaver}(1993)}]{Torres93}
{Torres-Dodgen}, A.~V. \& {Weaver}, W.~B. 1993, \pasp, 105, 693

\bibitem[{{Wang} \& {Chen}(2019)}]{Wang19}
{Wang}, S. \& {Chen}, X. 2019, \apj, 877, 116

\bibitem[{{Yao} {et~al.}(2018){Yao}, {Meyer}, {Covey}, {Tan}, \& {Da
  Rio}}]{Yao18}
{Yao}, Y., {Meyer}, M.~R., {Covey}, K.~R., {Tan}, J.~C., \& {Da Rio}, N. 2018,
  \apj, 869, 72

\bibitem[{{Zeilik} \& {Lada}(1978)}]{Zeilik78}
{Zeilik}, M., I. \& {Lada}, C.~J. 1978, \apj, 222, 896

\end{thebibliography}





\clearpage

\end{document}